\documentclass[aps, prl, superscriptaddress, twocolumn, letterpaper, 10pt, amsfonts, amsmath, amssymb]{revtex4-2}

\def\isarxiv{}             

\usepackage{graphicx}
\usepackage{hyperref} 
\usepackage{xcolor}
\usepackage{soul}
\hypersetup{
    colorlinks,
    linkcolor={blue!80!black},
    citecolor={blue!80!black},
    urlcolor={blue!80!black}
}
\usepackage{physics}
\usepackage{placeins}
\usepackage{bbold}

\usepackage{lipsum}
\usepackage{tikz}
\usepackage{subcaption}
\usepackage{ragged2e}

\usepackage[capitalize]{cleveref}
\crefformat{equation}{Eq.~(#2#1#3)}

\newcommand{\note}[1]{\textcolor{red}{[#1]}}

\makeatletter
\def\maketitle{
\@author@finish
\title@column\titleblock@produce
\suppressfloats[t]}
\makeatother

\begin{document}

\title{Adiabatic nonabelian braiding of imperfect Majoranas}

\author{Maximilian Nitsch}
\affiliation{Division of Solid State Physics and NanoLund, Lund University, Lund, Sweden}

\author{Viktor Svensson}
\affiliation{Department of Physics, University of Oslo, Oslo, Norway}

\author{William Samuelson}
\affiliation{Division of Solid State Physics and NanoLund, Lund University, Lund, Sweden}

\author{Konstantin Nestmann}
\affiliation{Division of Solid State Physics and NanoLund, Lund University, Lund, Sweden}

\author{Jeroen Danon}
\affiliation{Department of Physics, Norwegian University of Science and Technology, Trondheim, Norway}

\author{Karsten Flensberg}
\affiliation{Center for Quantum Devices, Niels Bohr Institute, University of Copenhagen, Copenhagen, Denmark}

\author{Rubén Seoane Souto}
\affiliation{Departamento de Física Teórica de la Materia Condensada, Condensed Matter Physics Center (IFIMAC), and Instituto Nicolás Cabrera, Universidad Autónoma de Madrid, Madrid, Spain}

\author{Martin Leijnse}
\affiliation{Division of Solid State Physics and NanoLund, Lund University, Lund, Sweden}

\date{\today}

\begin{abstract}
Demonstration of a nontrivial result of quasiparticle exchange (or braiding) is usually considered the definitive proof of a topological phase with nonabelian excitations, such as Majorana bound states (MBSs). However, in finite systems with disorder and smooth potential variations, the MBSs are imperfect in the sense that they are not fully isolated in space and can, to a varying degree, resemble conventional fermions. Here, we study the braiding properties of isolated MBSs, regular fermions, and anything in between. We find a way to compensate for the undesired splitting of the ground-state degeneracy which occurs during the protocol for imperfect MBS. This leads to a braiding outcome that depends on the degree of MBS isolation but remains robust and nonabelian except in the perfect fermion limit. Our protocol could be implemented in different platforms with nonabelian excitations, including quantum-dot-based minimal Kitaev chains.
\end{abstract}

\maketitle

\emph{Introduction.}
All elementary particles are either fermions or bosons, distinguished by whether particle exchange results in a sign change in the wavefunction or not. Quasiparticles in lower dimensions can in principle satisfy more general exchange rules. Nonabelian anyons are the most exotic example, with particle exchange (or braiding) described by a nontrivial operator acting on the multi-particle wavefunction~\cite{NayakReview2008}. Direct experimental demonstration of nonabelian braiding remains one of the main open challenges in modern physics~\cite{BeenakkerReview_20, Yazdani2023} and has potential applications in error-protected topological quantum computing~\cite{NayakReview2008,Sarma2015, Karzig_PRB2017, Marra2022}.

Majorana bound states (MBSs)~\cite{Wilczek2009, Alicea_RPP2012, Leijnse_Review2012, Beenakker_review2013}, zero-energy excitations in one-dimensional topological superconductors, arguably provide the experimentally most accessible platform to demonstrate nonabelian braiding (see, for example, Refs.~\cite{Ivanov2001,Alicea_NatPhys2011,Flensberg_PRL2011,Sau_PRB2011,van_Heck_NJP2012,Aasen_PRX2016,Hell2017,Plugge_NJP2017,Clarke2017Apr,BeenakkerReview_20}). Unfortunately, progress has been slowed by difficulties to unambiguously distinguish between MBSs and trivial states. 

The approach to realize MBSs by building up artificial Kitaev chains~\cite{Kitaev_2001} from quantum dots coupled via superconductors~\cite{Sau_NatComm2012, Leijnse_PRB2012, Fulga2013,Tsintzis2022,Liu_PRL22,Souto_chapter2024} has seen recent rapid experimental progress~\cite{Wang2022, Dvir2023, WangNatCom2023, Bordin_PRX23, vandriel2024, BordinPRL2024, tenhaafNature2024, Bording2025}. However, because these chains are typically rather short, these MBSs lack true topological protection~\cite{Leijnse_PRB2012}. Nonabelian braiding can still be possible with such MBSs, but requires fine-tuning~\cite{Tsintzis2024,Miles2025,Huang2024} and is susceptible to noise~\cite{Boross_2023}. In fact, the situation might be similar for supposedly topological realizations due to finite size effects, disorder, and smooth confining potentials~\cite{Vuik_SciPost19,Clarke_PRB2017,Prada_PRB2017,deng2016majorana,deng2018nonlocality}. In all these cases, the low-energy states can be described as MBSs that are imperfect in the sense that they are not fully isolated in space. On the other hand, a conventional zero-energy fermion can always be represented as two fully overlapping MBSs. Therefore, there is no strict border between nonabelian MBSs and conventional fermions, but rather a smooth transition between these two limiting cases. 

\begin{figure}[t]
    \setlength{\abovecaptionskip}{1pt}  
    \centering
    \begin{subfigure}[b]{0.48\columnwidth}
        \centering
        \begin{tikzpicture}
            \node[anchor=north west] (image) at (0,0) {\includegraphics[width=\linewidth]{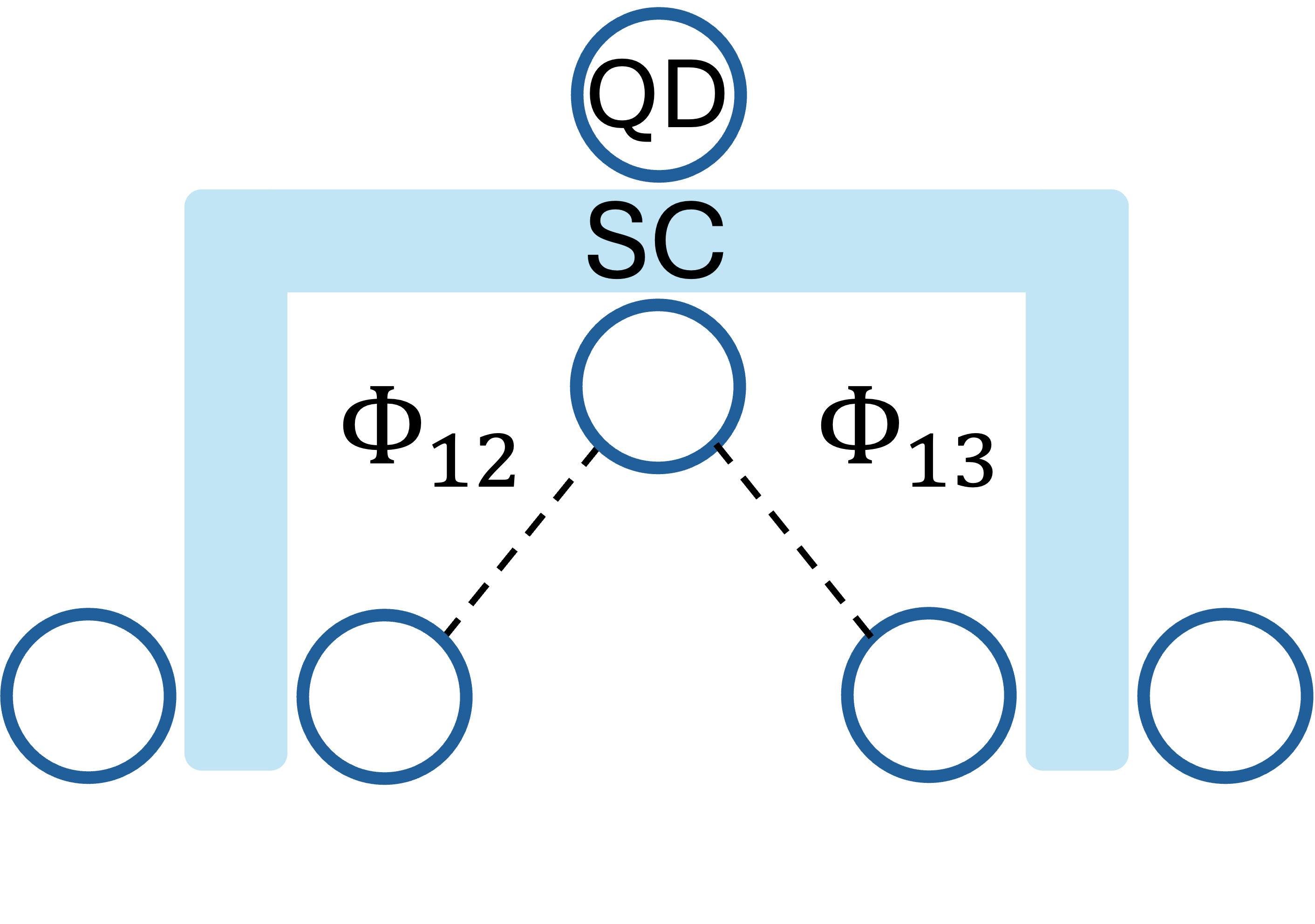}};
            \node[anchor=north west] at (image.north west) {\textbf{(a)}};
        \end{tikzpicture}
    \end{subfigure}
    \hfill
    \begin{subfigure}[b]{0.48\columnwidth}
        \centering
        \begin{tikzpicture}
            \node[anchor=north west] (image) at (0,0) {\includegraphics[width=\linewidth]{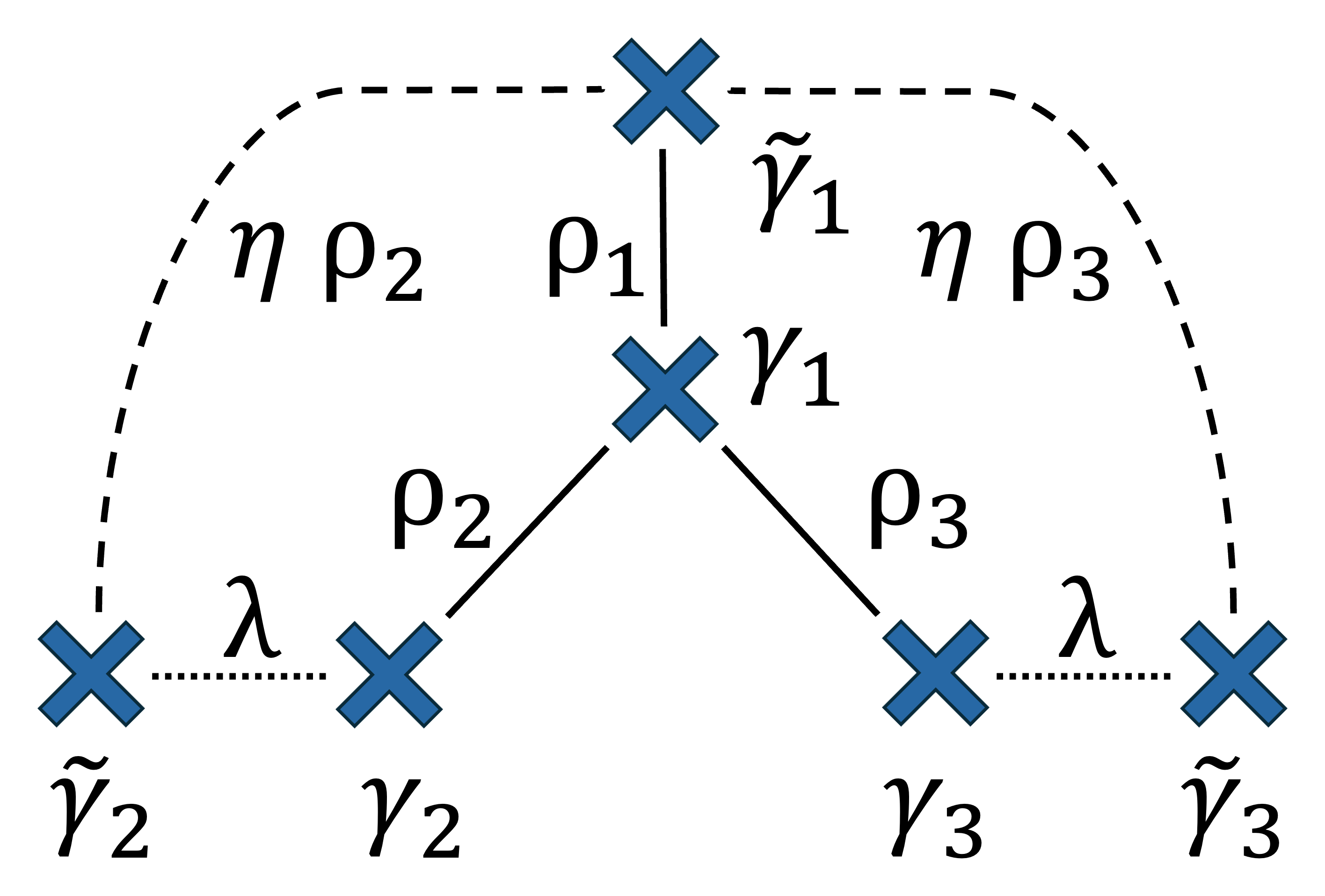}};
            \node[anchor=north west] at (image.north west) {\textbf{(b)}};
        \end{tikzpicture}
    \end{subfigure}

    \vspace{-1.0em}  

    \begin{subfigure}[b]{0.38\columnwidth}
        \centering
        \begin{tikzpicture}
            \node[anchor=north west] (image) at (0,0) {\includegraphics[width=\linewidth]{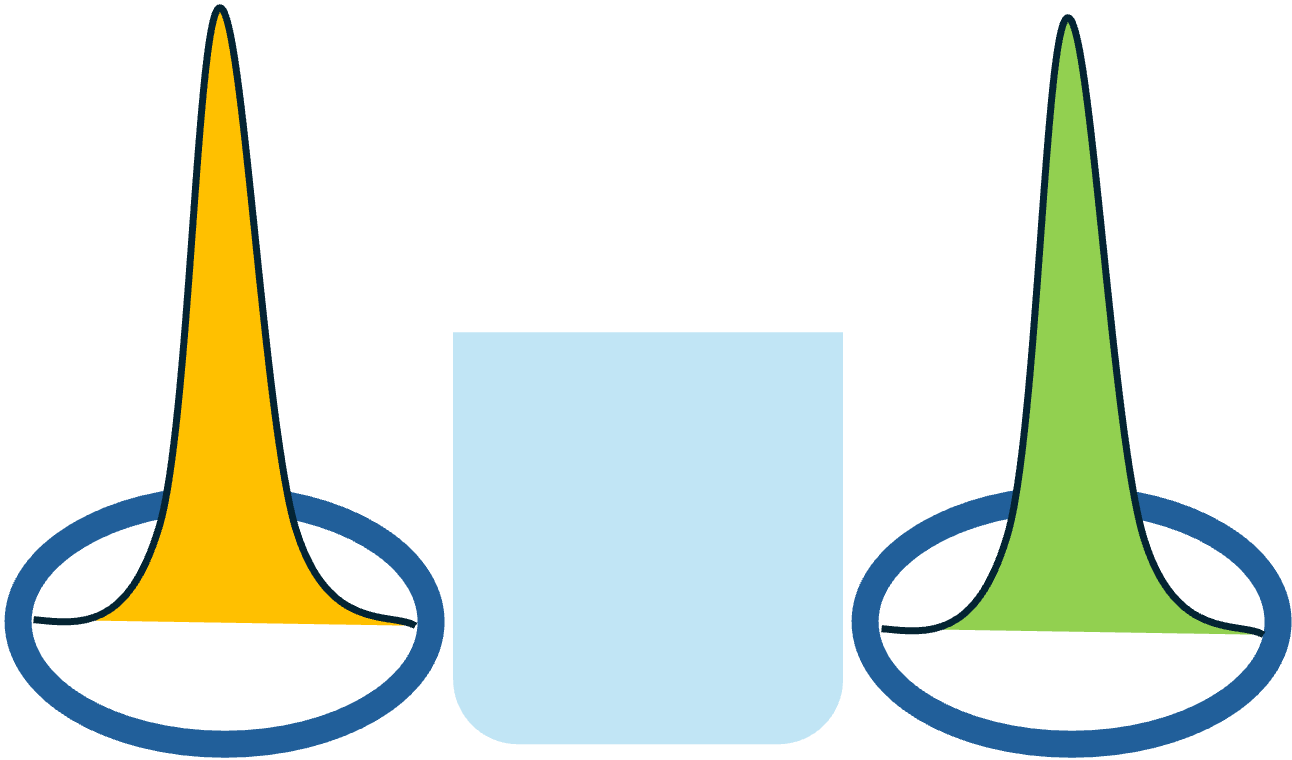}};
            \node[anchor=north west] at (image.north west) {\textbf{(c)}};
        \end{tikzpicture}
    \end{subfigure}
    \hfill
    \begin{subfigure}[b]{0.58\columnwidth}
        \centering
        \begin{tikzpicture}
            \node[anchor=north west] (image) at (0,0) {\includegraphics[width=\linewidth]{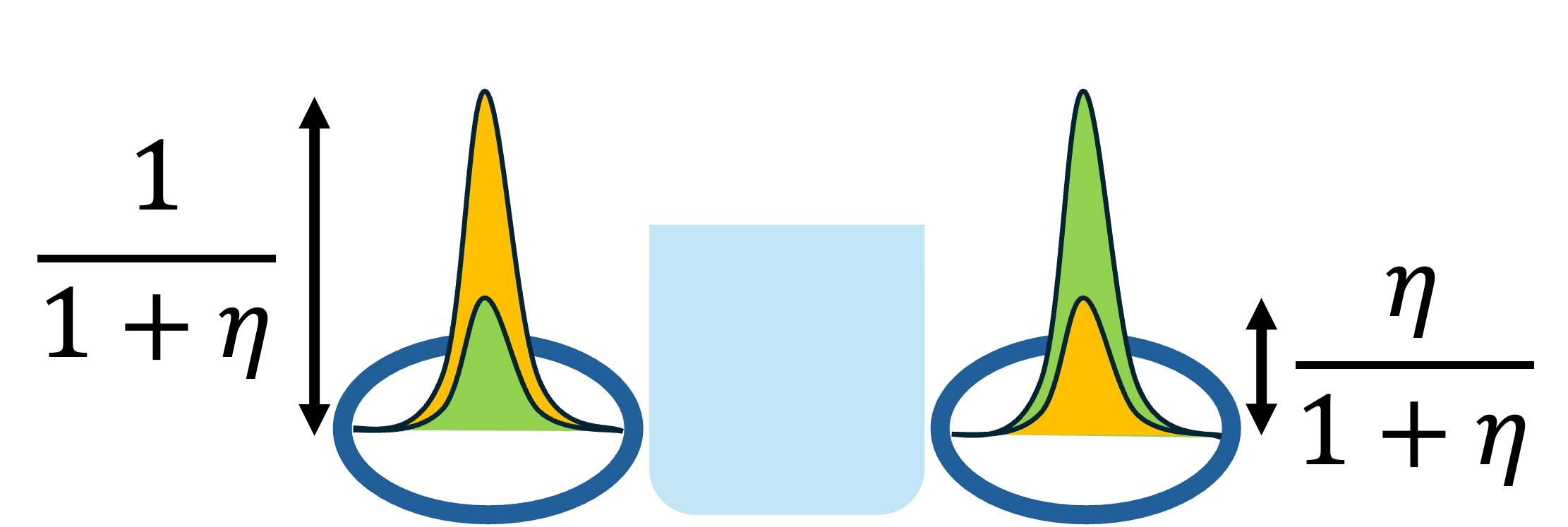}};
            \node[anchor=north west] at (image.north west) {\textbf{(d)}};
        \end{tikzpicture}
    \end{subfigure}

    \caption{\justifying
        (a) Example setup to implement coupling-based braiding in a system hosting MBSs in three minimal Kitaev chains based on quantum dots (QDs) connected via a superconductor (SC). Magnetic fluxes $\Phi_{12}, \Phi_{13}$ control the relative phases of the SC-QD couplings. 
        (b) Representation of the different MBS couplings used in  \cref{eq:Htot}.
        (c,d) Cartoon of isolated MBSs (c) and spatially overlapping MBSs (d).
    }
    \label{fig:combinedfigure}
\end{figure}
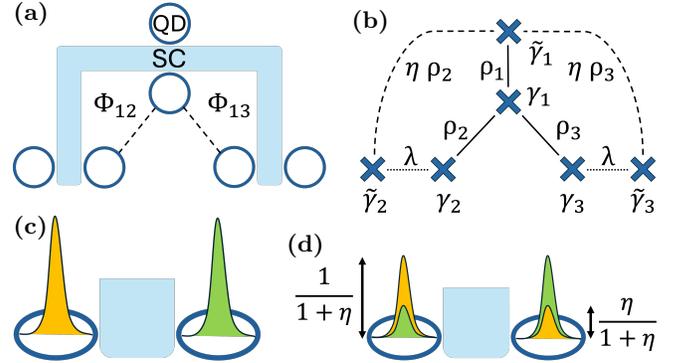

In this Letter, we investigate the braiding properties in a model that interpolates between the MBS and fermion limits. We focus on a specific braiding protocol based on cyclically coupling three systems~\cite{van_Heck_NJP2012,Sau_PRB2011,Karzig2015,Hell2017,Clarke2017Apr,Tsintzis2024,Miles2025}, one possible realization being three double quantum dots coupled via narrow superconducting regions as shown in~\autoref{fig:combinedfigure}(a). Each system is described by a model that hosts zero-energy MBSs that can be continuously tuned from being isolated (perfect MBS limit) to fully overlapping (conventional fermion limit). 

Even though each isolated system hosts a zero-energy state, deviations from fully isolated MBSs lead to a breaking of ground-state degeneracy when the systems are coupled during the protocol. Even a small induced ground-state splitting gives rise to a dynamical phase that destroys the stable braiding result, while a large splitting (compared to the inverse protocol time) forces the system to always stay in the unique global ground state and gives rise to a trivial braiding result. To reveal the fundamental adiabatic exchange properties of imperfect MBSs, we develop a way to compensate for this splitting to maintain perfect degeneracy throughout the braiding protocol. The resulting protocol performs a braiding operation that features a nonabelian result that is stable with respect to protocol speed and coupling strengths. Nevertheless, the result in general deviates from that of braiding isolated MBSs and converges to the trivial fermion result for fully overlapping MBSs. We present an analytical solution for the nonabelian Berry phase that determines the braiding result in the adiabatic and degenerate limit, and solve the general case numerically.

\emph{Model.} The braiding is realized  by connecting three systems, see \autoref{fig:combinedfigure}~(b), where couplings (lines) between different MBSs (crosses) are switched on and off~\cite{van_Heck_NJP2012,Sau_PRB2011,Karzig2015,Hell2017,Tsintzis2024,Miles2025}. A minimal representation is given by the Hamiltonian
\begin{equation}
\begin{aligned}
    \label{eq:Htot}
    H/\Delta = &\rho_1 \, i \gamma_1 \tilde{\gamma}_1 + \rho_2 \, i \gamma_1 \gamma_2 + \rho_3 \, i \gamma_1 \gamma_3 \\
    &+ \eta \, (\rho_2 \, i \tilde{\gamma}_1 \tilde{\gamma}_2 + \rho_3 \, i \tilde{\gamma}_1 \tilde{\gamma}_3) \\
    &+ \lambda \, (i \gamma_2 \tilde{\gamma}_2 + i \gamma_3 \tilde{\gamma}_3).
\end{aligned}
\end{equation}
The model includes three pairs of MBSs, $\gamma_k, \tilde{\gamma}_k$. Each pair encodes one fermionic state with two parity states $i \gamma_k \tilde{\gamma}_k= \pm 1$, thus yielding an eight-dimensional Hilbert space for the total system. Each term in \cref{eq:Htot} corresponds to a coupling between two MBSs (e.g. $\rho_1 \, i \gamma_1 \tilde{\gamma}_1$), assigning a certain energy ($\propto \pm \rho_1$) to each parity ($i \gamma_1 \tilde{\gamma}_1=\pm 1$). A graphical representation of all included couplings is shown in \autoref{fig:combinedfigure}~(b).
Total parity $\sigma = \prod_k i \gamma_k \tilde{\gamma}_k$ commutes with Eq.~$\eqref{eq:Htot}$ and is therefore conserved. One possible effective implementation of this Hamiltonian can be achieved via three minimal Kitaev chains~\cite{Leijnse_PRB2012} arranged in a T-junction, see \autoref{fig:combinedfigure}~(a) and Supplementary information (SI)~\cite{SI}.

The first line of \cref{eq:Htot} includes all desired couplings needed for the braiding protocol. These are the only couplings that occur if each system hosts MBSs that are fully isolated in space, see \autoref{fig:combinedfigure}~(c). Couplings between $\gamma_1$ and $\tilde{\gamma}_1, \gamma_2, \gamma_3$ open an energy gap $\Delta$ between the high- and low-energy sectors. This gap will be kept finite throughout the braiding protocol, which is guaranteed by enforcing the condition $\rho_1^2 + \rho_2^2 + \rho_3^2 = 1$. We thus obtain a well-defined ground state sector with two degenerate states in each total parity sector. 

We obtain the second line of \cref{eq:Htot} by allowing the outer MBSs $\tilde{\gamma}_k$ to partially overlap with the respective inner MBSs $\gamma_k$, see \autoref{fig:combinedfigure}~(d). Note that we focus on overlaps which do not introduce a finite energy associated with the parity $i \gamma_k \tilde{\gamma}_k$. The MBS overlap is parametrized by $\eta \in [0, 1]$, interpolating between isolated MBSs $(\eta=0)$ and trivial fermions $(\eta=1)$ where two MBSs are fully overlapping (in the region relevant for the coupling to the other systems). To facilitate an analytical solution below, we choose the same MBS overlap $\eta$ for all MBS pairs. The final result qualitatively agrees with a numerical solution for the general case with an effective $\eta$, see SI~\cite{SI}. A commonly used quality measure for MBSs is the Majorana polarization $M$~\cite{Sedlmayr2015,Sedlmayr2016,Aksenov2020,Tsintzis2022, Samuelson_PRB2024}. An increase in $\eta$ directly leads to a decrease of the Majorana polarization on the inner edges, $M=\frac{1-\eta}{1+\eta}$~\cite{SI}. 

In \cref{eq:Htot} there are no couplings between next-nearest neighbors ($\propto i \gamma_i \tilde{\gamma}_j$). We assume that these couplings have been turned off by an appropriate tuning of the phases of the couplings, which is always possible~\cite{Tsintzis2024}. In the setup in \autoref{fig:combinedfigure}(a) this is accomplished by the fluxes $\Phi_{12}$ and $\Phi_{13}$. There is, however, no general way to remove the remaining coupling terms in the second line of \cref{eq:Htot}, which will break the ground state degeneracy and prevent robust braiding operations~\cite{Tsintzis2024, Huang2024}. To compensate for this degeneracy breaking, we introduce couplings of the outer MBSs $\tilde{\gamma}_{2/3}$ to their inner neighbors $\gamma_{2/3}$, described by the third line of \cref{eq:Htot}. We choose these to be equal to simplify the analytical solution.

Four signs matter for the protocol performance: the total parity $\sigma$, the sign of $\rho_1$, and there could in principle be a "--" (rather than a "+") between the two terms in the second and third lines of \cref{eq:Htot}. A mismatch between the relative signs of lines two and three prevents establishing degeneracy via the correction. In the SI~\cite{SI} we show that the remaining eight sign combinations reduce to two types of protocols which differ from each other in the size of the gap and intermediate results during the protocol, but give the same final result for the double braid in the adiabatic limit, \cref{eq:fidelityAnalytical}. For the remainder of the paper, we take $\sigma = -1$ and all other signs positive.

\begin{figure}[t]
    \centering
    \includegraphics[width=1.0\linewidth]{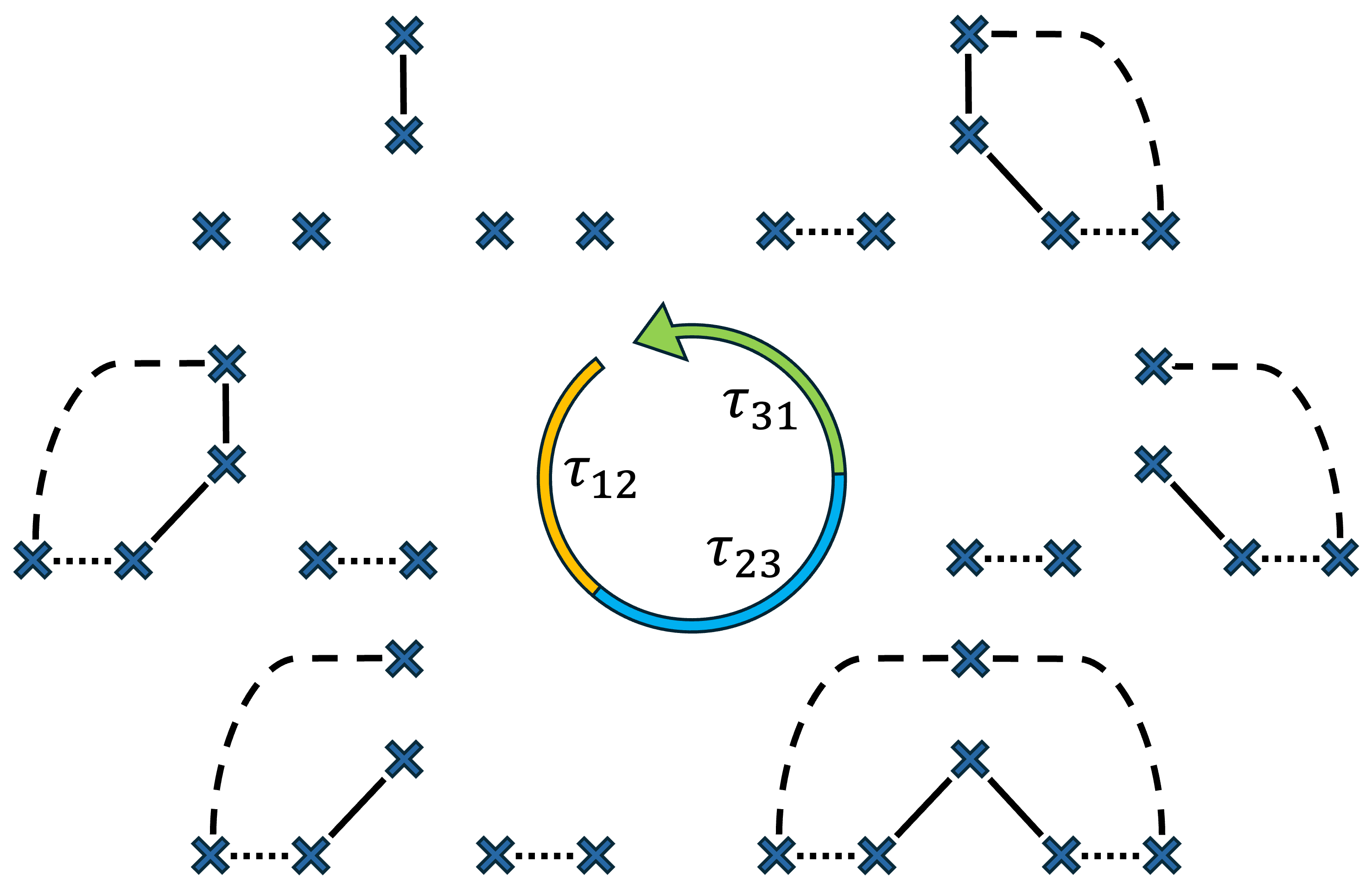}
    \caption{\justifying One cycle of the time-dependent tuning of the couplings in the braiding protocol. The protocol is split into three time intervals $\tau_{12}$ (orange), $\tau_{23}$ (blue), $\tau_{13}$ (green) connecting three coupling configurations as defined in \cref{eq:protocolSteps}.}
    \label{fig:protocol}
\end{figure}

\emph{Braiding protocol.} We start by considering perfectly isolated MBSs ($\eta=0$) without corrections ($\lambda=0$) and show that the coupling-based braiding protocol realizes an exchange of $\gamma_2$ and $\gamma_3$ by cyclically varying the different MBS couplings \cite{van_Heck_NJP2012,Sau_PRB2011,Karzig2015,Hell2017,Clarke2017Apr,Tsintzis2024,Miles2025}. This is done via a closed loop in the three-dimensional parameter space $\rho_1, \rho_2, \rho_3$, and we consider the limit where all parameters are changed  adiabatically slow compared to the gap $\Delta$, ensuring that the system stays in the ground state sector. The  loop in parameter space is split into three separate time intervals $\tau_{12}, \tau_{23}, \tau_{31}$
\begin{equation}
    \begin{aligned}
    \label{eq:protocolSteps}
        \tau_{12}: && \rho_1 = 1, \rho_2 = 0 && \rightarrow && \rho_1 = 0, \rho_2 = 1, \\
        \tau_{23}: && \rho_2 = 1, \rho_3 = 0 && \rightarrow && \rho_2 = 0, \rho_3 = 1, \\
        \tau_{31}: && \rho_3 = 1, \rho_1 = 0 && \rightarrow && \rho_3 = 0, \rho_1 = 1,
    \end{aligned}
\end{equation}
see the solid lines in \autoref{fig:protocol}.
Next will follow a detailed discussion of the first time interval, $\tau_{12}$, including a solution of the implemented rotation for a system hosting isolated MBSs. The other two protocol intervals $\tau_{23}, \tau_{31}$ work analogously.

We start by coupling MBSs $\gamma_1, \tilde{\gamma}_1$ ($\rho_1 = 1$), see the upper left panel of \autoref{fig:protocol}. The ground state then has $i\gamma_1 \tilde{\gamma}_1 = -1$ and, because of the four uncoupled MBSs, is two-fold degenerate within each total parity sector. In an experiment, the system should be initialized in one of these ground states (e.g., as described in Ref.~\cite{Tsintzis2024}). 

During the first time interval $\tau_{12}$, the coupling of $\gamma_1$ is transferred from $\tilde{\gamma}_1$ to $\gamma_2$ by first tuning $\rho_2 > 0$ and then $\rho_1 =0$, see the left column of \autoref{fig:protocol}. Tuning the couplings in this way guarantees a gap to excited states throughout the whole step. The exact parametrization of that exchange is unimportant as long as the changes are adiabatic.

In general, the result of the adiabatic braiding protocol can be found from the nonabelian Berry phase (see below). However, for $\eta = 0$ a simple intuitive argument suffices. During the time interval $\tau_{12}$, introducing the coupling to the second MBS, $\gamma_2$, effectively couples $\gamma_1$ to a superposition of $\tilde{\gamma}_1$ and $\gamma_2$ described by $\gamma_\Delta$
\begin{equation}
\begin{aligned}
    \label{eq:Ht12}
    H(\tau \in \tau_{12})/ \Delta  = \, i \gamma_1 \gamma_\Delta, && \gamma_\Delta = \rho_1 \tilde{\gamma}_1 + \rho_2 \gamma_2.
\end{aligned}
\end{equation}
The definition of $\gamma_\Delta$ required by the diagonalization directly yields an uncoupled MBS
\begin{equation}
    \label{eq:freeMajorana}
    \gamma_\Delta^\prime = \rho_1 \gamma_2 - \rho_2 \tilde{\gamma}_1,
\end{equation}
which is not part of the diagonalized Hamiltonian (\cref{eq:Ht12}). The above parametrization of $\gamma_\Delta^\prime$ is fully determined by demanding consistency with the initial conditions and anti-commutation relations
\begin{equation}
    \begin{aligned}
        \gamma_\Delta^\prime \stackrel{\rho_2 =0}{=} \gamma_2, && \{ \gamma_\Delta, \, \gamma_\Delta^\prime \} = 0.
    \end{aligned}
\end{equation}
Accordingly, the first step of the protocol described by \cref{eq:Ht12} and \cref{eq:freeMajorana} implements the transformation $\tilde{\gamma}_1 \rightarrow \gamma_2$, $\gamma_2 \rightarrow -\tilde{\gamma}_1$
while leaving all other operators unchanged. This corresponds to the unitary
\begin{equation}
    U_{12} = \exp(\frac{\pi}{4} \, \gamma_2 \tilde{\gamma}_1 ).
\end{equation}
The same kind of analysis also yields the results of the time intervals $\tau_{23}$ and $\tau_{31}$, where the coupling is transferred to $\gamma_3$ and finally back to $\tilde{\gamma}_1$. These steps are described by the unitaries
\begin{equation}
    \begin{aligned}
        U_{23} = \exp(\frac{\pi}{4} \, \gamma_3 \gamma_2), && U_{31} = \exp(\frac{\pi}{4} \, \tilde{\gamma}_1 \gamma_3).
    \end{aligned}
\end{equation}
Thus, the nonabelian Berry phase of the complete braiding protocol is
\begin{equation}
    U(\eta=0) = U_{31} \cdot U_{23} \cdot U_{12} = \exp(\frac{\pi}{4} \, \gamma_3 \gamma_2),
\end{equation}
implementing an exchange of $\gamma_2$ and $\gamma_3$.

Now we consider the case of imperfect MBSs, $\eta > 0$. This does not affect the initialization in the ground state of $i \gamma_1 \tilde{\gamma}_1$. However, turning on the coupling $\rho_2 \, i \gamma_1 \gamma_2$ now  introduces also couplings between $\tilde{\gamma}_1$ and $\tilde{\gamma}_2$, $\rho_2 \, \eta \, i \tilde{\gamma}_1 \tilde{\gamma}_2$ [dashed lines in Figs.~\ref{fig:combinedfigure}(b) and \ref{fig:protocol}], lifting the ground state degeneracy. This degeneracy is restored via our correction term $\lambda \, (i \gamma_2 \tilde{\gamma}_2 + i \gamma_3 \tilde{\gamma}_3)$ [dotted lines in Figs.~\ref{fig:combinedfigure}(b) and \ref{fig:protocol}], which has to be tuned at the same time as $\rho_1, \rho_2$. The exact relation between $\lambda$ and $\rho_1, \rho_2$ needed to maintain ground-state degeneracy will be established below.

We can use the intuition from the $\eta = 0$ result to guess the possible braiding results for the general case. The protocol exchanges MBSs with each other when they are combined into a superposition of the ground-state basis during a protocol interval, like $\gamma_\Delta, \gamma_\Delta^\prime$. This diagonalization always combines pairs of next-nearest neighbors that couple to a shared MBS. Therefore, we can divide all available MBSs into two groups: $\gamma_2, \gamma_3, \tilde{\gamma}_1$ and $\tilde{\gamma}_2, \tilde{\gamma}_3, \gamma_1$. The couplings between the groups can be arbitrary, but there are no direct couplings between two MBSs of the same group. We find that rotations can only consist of exchanges within those groups \cite{SI}. Combinations including $\gamma_1, \tilde{\gamma}_1$ can be excluded because they correspond to rotations to excited states of the initial and final configuration $(\rho_1=1)$. Therefore, the most general operator the protocol can implement must be of the form 
\begin{equation}
    \label{eq:qualtitativeResult}
    U(\eta) = \exp(\phi \, \gamma_3 \gamma_2) \, \exp(\tilde{\phi} \, \tilde{\gamma}_3 \tilde{\gamma}_2).
\end{equation}
Next, we present an analytical solution for $\phi, \tilde{\phi}$, valid for arbitrary $\eta$. This solution is split into two steps. First, we diagonalize the Hamiltonian. This enables us to establish degeneracy by tuning $\lambda$ and to describe the ground-state sector via a projector. This projector is then used to find the time evolution in the adiabatic limit, which is described by the nonabelian Berry phase.

\emph{Identifying the ground-state MBSs.}
We start by rewriting the diagonalization of $H$ for isolated MBSs introduced in \cref{eq:Ht12} which was valid for the protocol interval $\tau_{12}$. We extend this over the whole protocol time
by redefining $\gamma_\Delta = \rho_1 \, \tilde{\gamma}_1 + \rho_2 \, \gamma_2 +  \rho_3 \, \gamma_3$, similar to $\cite{Karzig2015}$. $\gamma_\Delta$ constitutes the MBS that establishes the gap $\Delta$ with $\gamma_1$ throughout the protocol, thereby defining the ground-state sector.

Due to the MBS overlap $\eta$ and correction terms $\lambda$, we need to take additional steps to diagonalize the full system of \cref{eq:Htot}, see SI~\cite{SI}. From this diagonalization, we obtain two pieces of information that are necessary to implement the correction and to describe the effective ground-state sector. First, we find the MBSs $\gamma_1^D$ and $\gamma_\Delta^D$, which together define the ground-state sector via $i \gamma_1^D \gamma_\Delta^D=-1$, while taking errors and corrections into account. This introduces the ground state projection operator
\begin{equation}
    \label{eq:projOp}
    P = \frac{1}{2}(1- i \gamma_1^D \gamma_\Delta^D).
\end{equation}
Furthermore, $\eta$ introduces a finite splitting $\varepsilon$ of the ground state sector. We compensate for this splitting by tuning $\lambda$. From the diagonalized Hamiltonian, we find the degeneracy condition
\begin{equation}
\label{eq:degeneracyCondition}
    \varepsilon (\lambda, \eta) - \lambda = 0,
\end{equation}
which defines the non-linear dependence of the correction $\lambda$ on the MBS overlap $\eta$, which is furthermore dependent on the protocol parameters $\rho_1, \rho_2, \rho_3$ and therefore time-dependent throughout the protocol. The required correction depends on $\sigma$ \cite{SI}, which prevents using this protocol for actual topological quantum computing, where we need to perform gates on the qubits without knowing their state. There is no obvious, general analytical solution to \cref{eq:degeneracyCondition}, but we find in the calculation of the Berry phase that it only requires the solution of this expression in the second protocol interval $\tau_{23}$, defined by $\rho_1=0$, where it simplifies crucially, see \cite{SI}.

\emph{Calculating the Berry phase.}
As the next step, we use \cref{eq:projOp} to obtain the braiding operator in the ground-state sector from the nonabelian Berry phase \cite{Avron_CommMatPhys1989, hassler2024topologicalquantumcomputing}
\begin{equation}
    U(\eta) = \mathcal{T} \exp\left(- \oint  [P, \partial_\omega P] \, d \omega\right),
\end{equation}
where $\mathcal{T}$ is the time-ordered path integral. We simplify the calculation by splitting it into the three protocol intervals defined in \cref{eq:protocolSteps} $U(\eta) = U_{31} \cdot U_{23} \cdot U_{12}$, and solve each interval separately, see \cite{SI}. The solution yields the analytical expressions for the angles $\phi(\eta), \tilde{\phi}(\eta)$ in~\cref{eq:qualtitativeResult}
\begin{equation}
    \begin{aligned}\label{eq:lambdaphi}
        \phi = \frac{\pi}{4} \left( 1 -\frac{ \eta^2}{\sqrt{1+\eta^2 + \eta^4}} \right), \\
        \tilde{\phi} = \frac{\pi}{4} \left( 1 - \frac{1}{\sqrt{1+\eta^2 + \eta^4}} \right).
    \end{aligned}
\end{equation}
Note that these angles are only correct when acting with $U(\eta)$ on the ground-state sector.

\emph{Discussion.}
The result of the braiding protocol in the fully adiabatic limit for arbitrary MBS overlaps $\eta$ and perfect corrections $\lambda$ is given by~\cref{eq:qualtitativeResult} together with \cref{eq:lambdaphi}. We can further simplify the solution using conservation of total parity, $\tilde{\gamma}_3 \tilde{\gamma}_2 = - \, \gamma_3 \gamma_2$, 
yielding
\begin{equation}
    \label{eq:Ueff}
    U(\eta) = \exp\left( (\phi - \tilde{\phi}) \, \gamma_3 \gamma_2\right).
\end{equation}
This representation highlights that the braiding protocol implements a partial exchange of $\gamma_2, \gamma_3$ given by the angle $\phi_\text{eff}(\eta) = \phi - \tilde{\phi}$, 
interpolating between a full exchange for isolated MBSs $\phi_\text{eff}(\eta = 0) = \frac{\pi}{4}$ and a trivial exchange for a fermionic system $\phi_\text{eff}(\eta = 1) = 0$. 
\begin{figure}[t]
    \centering
    \includegraphics[width=1.0\linewidth]{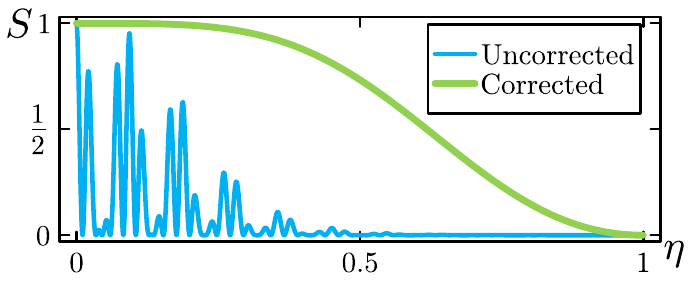}
    \caption{\justifying MBS similarity $S$ for a double braid as a function of the MBS overlap $\eta$. Green line: analytical result given by \cref{eq:fidelityAnalytical} for a perfectly corrected protocol. Blue oscillating line: Numerical result for the uncorrected protocol (details in Sect. VII. in the SI~\cite{SI}).}
    \label{fig:quality}
\end{figure}

Next, we introduce the MBS similarity $S(U)$ that quantifies how much the braiding result resembles that of fully isolated MBSs. We base $S(U)$ on a double braid, i.e., on $U^2$, which provides a clearer and more sensitive experimental signature than a single braid~\cite{Clarke2017Apr,Tsintzis2024}. From \cref{eq:Ueff} we find the simple result
\begin{equation}
    \label{eq:fidelityAnalytical}
    S(U) = \Tr{U^2(\eta)^{\dagger} \, U^2(0)}^2/4 = \sin^2\left(\frac{\pi}{2}\sqrt{\frac{(1-\eta^2)^3}{(1-\eta^6)}}\right),
\end{equation}
which quantifies the coupling between the unitary transformation $U^2(\eta)$ implemented by the double braid protocol and the expected result for an isolated MBS system, $U^2(\eta=0) = \exp\left(\frac{\pi}{2} \, \gamma_3 \gamma_2\right)$, see green line in~\autoref{fig:quality}. For $\eta \lesssim 0.25$, the result very closely resembles the braiding of isolated MBSs. For larger values, the fermionic character becomes increasingly evident and the result approaches the trivial fermion result $U^2(\eta=1) = \mathbb{1}$. However, the result remains nonabelian for all $\eta < 1$. 

We compare this to the result of the braiding protocol without corrections ($\lambda = 0$), which leads to the transformation $U_\text{uncorr}$, suffering from breaking of the ground-state degeneracy. The result is shown in \autoref{fig:quality} (blue line), which is based on numerically solving the time-dependent Schrödinger equation, see Sect. VII. in the SI \cite{SI} for a description of the method as well as the parameters used and Ref. \cite{nitsch_code_braiding} for the code. The uncorrected curve shows MBS behavior in a very small region around $\eta \approx 0$. For larger MBS overlaps, the dynamical phase leads to an additional rotation within the almost degenerate ground state sector \cite{Tsintzis2024, Huang2024}. For even larger MBS overlaps, $\eta \gtrsim 0.25$, the energy splitting within the ground state sector becomes so large that the protocol acts adiabatically concerning the splitting and the system remains in the initial state. While these three regimes are universal, the position of boundaries between them along $\eta$ depends on the detailed protocol parametrization and relation between the induced energy splitting and the speed of parameter changes.

$S(U)$ is not easily accessible in experiments. However, it is closely related to the result of an experiment where: $(i)$ the MBS pairs $\gamma_2, \tilde{\gamma}_2$ and $\gamma_3, \tilde{\gamma}_3$ are  initialized in a given parity state; $(ii)$ the double braid is performed; $(iii)$ the parities of the same pairs are measured again. The result for fully isolated MBS is a parity flip, while the trivial fermion result is no change compared to the initialized parities. 

\emph{Conclusions.}
In this Letter, we have presented a generalized coupling-based braiding protocol applied to a model that interpolates between isolated MBSs and conventional zero-energy fermions, the latter being equivalent to spatially fully overlapping MBSs. The MBS overlaps introduce a ground state splitting during the protocol, which we showed can be corrected for by additional MBS couplings. Except in the perfect fermion limit, the braiding result is nonabelian and robust to changes in protocol details such as the strength and duration of MBS couplings. Our work expands the notion of what constitutes a nonabelian state, with possible impacts ranging from topological quantum computing protocols to classification and experimental detection of topological phases. We believe that quantum-dot-based minimal Kitaev chains are an ideal platform to realize our protocol, as they allow control of all required parameters.



\emph{Acknowledgements}
We acknowledge funding from the European Research Council (ERC) under the European Unions Horizon 2020 research and innovation programme under Grant Agreement No. 856526, the Spanish CM “Talento Program” (project No. 2022-T1/IND-24070), the Spanish Ministry of Science, innovation, and Universities through Grant No. PID2022-140552NA-I00, the Swedish Research Council under Grant Agreement No. 2020-03412, the European Union’s Horizon Europe research and innovation programme under the Marie Skłodowska-Curie grant agreement No. 101126636, and NanoLund.

\bibliography{main}

\ifdefined\isarxiv
    \clearpage
    \onecolumngrid
    \ifdefined\isarxiv
    \setcounter{equation}{0}     
    \setcounter{figure}{0}       
\else
    \documentclass[aps, prl, superscriptaddress, twocolumn, letterpaper, 10pt, amsfonts, amsmath, amssymb]{revtex4-2}
    
    \usepackage{graphicx}
    \usepackage{hyperref} 
    \usepackage{xcolor}
    \usepackage{ragged2e} 
    \usepackage{soul}
    \hypersetup{
        colorlinks,
        linkcolor={blue!80!black},
        citecolor={blue!80!black},
        urlcolor={blue!80!black}
    }
    \usepackage{physics}
    \bibliographystyle{apsrev4-2}
    
    \usepackage{placeins}

    \usepackage{lipsum}
    \usepackage{tikz}
    \usepackage{subcaption} 
    \usepackage{dsfont}
    \usepackage{bbold}
    
    \usepackage[capitalize]{cleveref}
    \crefformat{equation}{Eq.~(#2#1#3)}

    \newcommand{\note}[1]{\textcolor{red}{[#1]}}

    \setlength{\belowcaptionskip}{-10pt}
    
    \newcommand{\viktor}[1]{{\color{purple}{#1}}}
    
    \makeatletter
    \def\maketitle{
    \@author@finish
    \title@column\titleblock@produce
    \suppressfloats[t]}
    \makeatother
    
    \begin{document}
    
    \author{Maximilian Nitsch, Viktor Svensson, William Samuelson, Konstantin Nestmann, \\ Jeroen Danon, Karsten Flensberg, Rubén Seoane Souto, Martin Leijnse}
\fi

\setcounter{secnumdepth}{2}

\renewcommand{\theequation}{S\arabic{equation}}
\renewcommand{\thefigure}{S\arabic{figure}}

\date{\today}

\title{Supplementary information for Adiabatic nonabelian braiding of imperfect Majoranas}
\maketitle

\section{Realization of the braiding protocol via Minimal Kitaev chains}
Here we describe a possible realization of the generalized braiding protocol based on minimal Kitaev chains~\cite{Kitaev_2001, Sau_NatComm2012, Leijnse_PRB2012, Fulga2013,Tsintzis2022,Liu_PRL22,Souto_chapter2024}. A single such system consists of two spin-polarized quantum dots ($l=1,2$) holding a single fermionic state each, that couple via a narrow superconductor
\begin{equation}
    H = \sum_{l=1,2} \xi_l \, n_l+t_\text{cot} \, c_1^\dagger c_2 + \Delta_\text{car} c_1^\dagger c_2^\dagger + {\rm H.c.}.
\end{equation}
Here, $t_{\rm cot}$ and $\Delta_{\rm car}$ are the elastic co-tunneling and crossed Andreev reflection amplitudes, which are tunable using electrostatic gates~\cite{Liu_PRL22,Dvir2023,Bordin_PRX23}. For $t_\text{cot} = \Delta_\text{car}, \, \xi_1, \xi_2 =0$ this system has a degenerate ground state sector described by two MBSs $\gamma, \tilde{\gamma}$ each localized on one dot, see Ref.\,\cite{Leijnse_PRB2012}. Deviations from $t_\text{cot} = \Delta_\text{car}$ lead to a splitting of the ground state expressed via a coupling $\epsilon$ of the two MBSs
\begin{equation}
\begin{aligned}
    H_\text{c} = \epsilon \, i \gamma \tilde{\gamma}, && \epsilon = \Delta_\text{car} - t_\text{cot}.
\end{aligned}
\end{equation}
Next, we build a braiding setup by connecting three minimal Kitaev chains, as shown in Fig. 1(a) of the main paper. Each system hosts a pair of MBSs. The inner dots of the three systems are connected via tunnel couplings
\begin{equation}
    \label{eq:fullPMMHam}
    H = \sum_{k=1}^3 \epsilon_k \, i \gamma_k \tilde{\gamma}_k + t_{12} \, i \gamma_1 \gamma_2 + t_{13} \, i \gamma_1 \gamma_3,
\end{equation}
where fluxes $\Phi_{12}, \Phi_{13}$ through the superconducting loops have been used to make the tunnel couplings $t_{12}, t_{13}$ real-valued.

The braiding protocol is defined by the Hamiltonian in \cref{eq:fullPMMHam} where the time-dependent parameters $\epsilon_1(t), t_{12}(t), t_{13}(t)$ implement the coupling-based braiding protocol~\cite{van_Heck_NJP2012,Sau_PRB2011,Karzig2015,Hell2017,Clarke2017Apr,Tsintzis2024,Miles2025} while we keep $\epsilon_2, \epsilon_3=0$. 

Next, we generalize the system by considering the possibility of spatially overlapping MBSs, meaning that the 'outer' MBSs $\tilde{\gamma}_k$ are partially located on the inner dot together with $\gamma_k$. This leads to a change in the tunnel coupling between the inner dots, which is described by the transformation
\begin{equation}
\label{eq:AndreevTransform}
    \gamma_k \rightarrow \frac{1}{\sqrt{1+ \zeta^2}} \, \gamma_k + i \frac{\zeta}{\sqrt{1+ \zeta^2}} \, \tilde{\gamma}_k,
\end{equation}
that interpolates between perfectly localized MBSs ($\zeta=0$) and trivial states ($\zeta=1$). For the two-site Kitaev chain, this parameter directly connects to the MBS polarization $M$ via $M = \frac{1-\zeta^2}{1+\zeta^2}$. In the case of minimal Kitaev chains, this introduces a dependence of $\zeta$ on the energy detuning $\epsilon_k$. However, in the analysis in the main paper, we assume that $\zeta$ is a constant system parameter independent of the variables $\epsilon_k$.

The transformation in \cref{eq:AndreevTransform} changes the system Hamiltonian as
\begin{equation}
    H \rightarrow H + \tilde{H},
\end{equation}
which rescales the tunnel coupling $t_{12}, t_{13}$ in $H$ and introduces contributions via $\tilde{H}$ that stem from the fermionic character of the coupled operators
\begin{equation}
\begin{aligned}
    H &= \epsilon_1 \, i \gamma_1 \tilde{\gamma}_1 
            + \frac{1}{1+\zeta^2} \, t_{12} \, i \gamma_1 \gamma_2
            + \frac{1}{1+\zeta^2}\, t_{13} \, i \gamma_1 \gamma_3, \\
    \tilde{H} &= \frac{\zeta^2}{1+\zeta^2} \, t_{12} \, i \tilde{\gamma}_1 \tilde{\gamma}_2 + \frac{\zeta^2}{1+\zeta^2} \, t_{13} \, i \tilde{\gamma}_1 \tilde{\gamma}_3.
\end{aligned}
\end{equation}
One of the main effects of $\tilde{H}$ is to lift the ground state degeneracy $\delta E \neq 0$~\cite{Tsintzis2024, Huang2024}. If the protocol is adiabatic compared to $\delta E$, its outcome is trivial: the system remains in its original state. For a protocol that is diabatic with respect to $\delta E$, the splitting instead leads to a dynamical phase, which makes the result dependent on the exact strengths and durations of the time-dependent couplings.

The ground state degeneracy can be recovered via the following correction term
\begin{equation}
    H_\text{corr} = \Lambda \, i \gamma_2 \tilde{\gamma}_2 + \Lambda \, i \gamma_3 \tilde{\gamma}_3,
\end{equation}
which uses the two remaining MBS overlaps in the left and right MBS pairs, which for a minimal Kitaev chain can be tuned via $t_\text{cot}$ and $\Delta_\text{car}$. In general, $\Lambda(t)$ is a time-dependent parameter that has to be tuned along with the protocol. Choosing the same overlap for both systems is a possible, but not a necessary, choice that simplifies the analytical solution. This yields the full Hamiltonian.
\begin{equation}
\label{eq:fullHam}
    H_\text{tot} = H + \tilde{H} + H_\text{corr}\,.
\end{equation}
The energy splitting $\Delta$ between the degenerate ground state sector and the excited states is given by
\begin{equation}
\label{eq:gap}
    \Delta = \sqrt{\epsilon_1^2 + \frac{1}{(1+\zeta^2)^2} \, \left(t_{12}^2 + t_{13}^2\right)},
\end{equation}
which we use to define the dimensionless parameters
\begin{equation}
\begin{aligned}
    \rho_1 &= \epsilon_1/\Delta, && \rho_{(k=2,3)} = \frac{1}{1+\zeta^2} \, t_{1 k}/\Delta, \\
    \lambda &= \Lambda/ \Delta, && \eta = \zeta^2.
\end{aligned}
\end{equation}
We now arrive at the dimensionless Hamiltonian in Eq.~(1) of the main paper
\begin{equation}
\begin{aligned}
    \label{eq:hDimless}
    h = H_\text{tot}/\Delta= \, i \gamma_1 \gamma_\Delta
    &+ \eta \, (\rho_2 \, i \Tilde{\gamma}_1 \Tilde{\gamma}_2 + \rho_3 \, i \Tilde{\gamma}_1 \Tilde{\gamma}_3) \\
    &+ \lambda \, (i \gamma_2 \Tilde{\gamma}_2 + i \gamma_3 \Tilde{\gamma}_3).
\end{aligned}
\end{equation}

\section{Diagonalizing the Hamiltonian}
In the following we will to through the diagonalization of \cref{eq:hDimless}, which is the first step needed in the analytical solution for the result of the adiabatic braiding protocol. We start by introducing $\gamma_\Delta$ which is defined by
\begin{equation}
    \label{eq:deltaMajorana}
    \gamma_\Delta = \cos \theta \, \Tilde{\gamma}_1 + \sin \theta \, \cos \phi \, \gamma_2 +  \sin \theta \, \sin \phi \, \gamma_3,
\end{equation}
expressed in via the angles $\theta, \phi$ defined as
\begin{equation}
    \begin{aligned}
        \rho_1 &= \cos(\theta), \\
        \rho_2 &= \sin(\theta) \cos(\phi), \\
        \rho_3 &= \sin(\theta) \sin(\phi).
    \end{aligned}
\end{equation}
This rotation of the $\Tilde{\gamma}_1, \gamma_2, \gamma_3$ basis enforces the introduction of the following two rotated orthogonal MBSs
\begin{equation}
    \begin{aligned}
    \label{eq:angleMajoranas}
        \gamma_\theta^\prime &= -\sin \theta \, \Tilde{\gamma}_1 + \cos \theta \, \cos \phi \, \gamma_2 +  \cos \theta \, \sin \phi \, \gamma_3, \\
        \gamma_\phi^\prime &= - \sin \phi \, \gamma_2 + \cos \phi \, \gamma_3.
    \end{aligned}
\end{equation}
In a similar way, we choose to rotate the MBSs $\Tilde{\gamma}_2, \Tilde{\gamma}_3$ as
\begin{equation}
    \begin{aligned}
    \label{eq:tildeMajoranas}
        \gamma_\eta &= \cos \phi \, \Tilde{\gamma}_2 + \sin \phi \, \Tilde{\gamma}_3, \\
        \gamma_\eta^\prime &= -\sin \phi \, \Tilde{\gamma}_2 + \cos \phi \, \Tilde{\gamma}_3.
    \end{aligned}
\end{equation}
$\gamma_1$ is the only MBS that we do not rotate. Rewriting the Hamiltonian in this rotated basis yields
\begin{equation}
\begin{aligned}
    h = i \gamma_1 \gamma_\Delta 
    &+ i \, \gamma_\Delta \gamma_\eta \, (\lambda \sin \theta + \eta \cos \theta \sin \theta) \\
    &+ i \, \gamma_\eta \gamma_\theta^\prime \, (\eta \sin^2 \theta - \lambda \cos \theta ) \\
    &+ i \, \gamma_\phi^\prime \gamma_\eta^\prime \, \lambda.
\end{aligned}
\end{equation}
What we achieved by these rotations is to decouple the MBS pair $ \gamma_\phi^\prime, \gamma_\eta^\prime$. Now we aim to diagonalize the remaining Hamiltonian by the ansatz
\begin{equation}
\begin{aligned}
\label{eq:diagonalMajoranas}
    \gamma_1^D &= \alpha \, \gamma_1 + \beta \, \gamma_\eta, && 
    \gamma_\eta^D = \alpha \, \gamma_\eta - \beta \, \gamma_1, \\
    \gamma_\Delta^D &= \mu \, \gamma_\Delta + \nu \, \gamma_\theta^\prime &&
    \gamma_\theta^{\prime \, D} = \mu \, \gamma_\theta^\prime - \nu \, \gamma_\Delta.
\end{aligned}
\end{equation}
Inserting this ansatz in the Hamiltonian yields
\begin{equation}
    \begin{aligned}
        h =& i \gamma_1^D \gamma_\Delta^D (\alpha \mu - \Tilde{\lambda} \beta \mu + \Tilde{\eta} \beta \nu) \\
        &+ i \gamma_\eta^D \gamma_\theta^{\prime D} (\beta \nu + \Tilde{\lambda} \alpha \nu + \Tilde{\eta} \alpha \mu) \\
        &+ i \gamma_\Delta^D \gamma_\eta^D (\mu \beta + \Tilde{\lambda} \alpha \mu - \Tilde{\nu} \alpha \nu) \\
        &+ i \gamma_1^D \gamma_\theta^{\prime D} ( - \alpha \nu + \Tilde{\lambda} \beta \nu + \Tilde{\nu} \beta \mu) \\
        &+ i \gamma_\phi^\prime \gamma_\eta^\prime \, \lambda,
    \end{aligned}
\end{equation}
where we defined
\begin{equation}
    \begin{aligned}
        \Tilde{\lambda} &= \lambda \sin \theta + \eta \cos(\theta) \sin(\theta), \\
        \Tilde{\eta} &= \eta \sin ^2 \theta - \lambda \cos \theta.
    \end{aligned}
\end{equation}
Next, we demand that this new basis decouples the MBS pair $\gamma_1^D, \gamma_\Delta^D$ from the pair $ \gamma_\eta^D,\gamma_\theta^{\prime D}$, which means that the terms proportional to $i \gamma_\Delta^D \gamma_\eta^D$ and $i \gamma_1^D \gamma_\theta^{\prime D}$ have to be zero
\begin{equation}
    \begin{aligned}
    \label{eq:decouplingMajs}
        \mu \beta + \Tilde{\lambda} \alpha \mu - \Tilde{\nu} \alpha \nu \stackrel{!}{=} 0, \\
        - \alpha \nu + \Tilde{\lambda} \beta \nu + \Tilde{\nu} \beta \mu \stackrel{!}{=} 0.
    \end{aligned}
\end{equation}
This can be satisfied by introducing two angles $\theta_\alpha, \theta_\mu$ defined as
\begin{equation}
\label{eq:thetaSolutions}
    \begin{aligned}
        \theta_\mu &= - \frac{1}{2} \arctan(\frac{2 \Tilde{\lambda} \Tilde{\eta}}{1 + \Tilde{\lambda}^2 - \Tilde{\eta}^2} ), \\
        \theta_\alpha &= -\arctan(- \Tilde{\eta} \tan \theta_\mu + \Tilde{\lambda}),
        \end{aligned}
\end{equation}
which yield the previously introduced basis components 
\begin{equation}
    \begin{aligned}
    \label{eq:angleParametrization}
        \alpha = \cos \theta_\alpha, && \beta = \sin \theta_\alpha, \\
        \mu = \cos \theta_\mu, && \nu = \sin \theta_\mu,
    \end{aligned}
\end{equation}
and allows writing the Hamiltonian in the fully diagonal form
\begin{equation}
        \begin{aligned}
        h = \Tilde{\Delta} \, i \gamma_1^D \gamma_\Delta^D
        + \varepsilon \, i \gamma_\eta^D \gamma_\theta^{\prime D}
        + \lambda \, i \gamma_\phi^\prime \gamma_\eta^\prime,
    \end{aligned}
\end{equation}
The Hamiltonian is defined such that the overall gap is given by 
\begin{equation}
    \Tilde{\Delta} = \alpha \mu - \Tilde{\lambda} \beta \mu + \Tilde{\eta} \beta \nu,
\end{equation}
that can be shown to be positive. This means that the low-energy sector is defined by
\begin{equation}
    \label{eq:groundStateCond}
    i \gamma_1^D \gamma_\Delta^D = -1,
\end{equation}
which will stay constant throughout the protocol. Within this low-energy sector there are the additional smaller splittings $\lambda$ and
\begin{equation}
    \label{eq:splitting}
    \varepsilon = \beta \nu + \Tilde{\lambda} \alpha \nu + \Tilde{\eta} \alpha \mu = \Tilde{\eta} \, \alpha/ \mu,
\end{equation}
coupled to the parities $i \gamma_\phi^\prime \gamma_\eta^\prime$ and $i \gamma_\eta^D \gamma_\theta^{\prime D}$. The last step in \cref{eq:splitting} can be shown via \cref{eq:decouplingMajs}. In general, this means that the low-energy states are not degenerate. Finally, we determine how to tune $\lambda$ to establish degeneracy. For this, we make use of the conservation of total parity. The basis definitions Eqs. (\cref{eq:deltaMajorana}), (\cref{eq:angleMajoranas}), (\cref{eq:tildeMajoranas}), and (\cref{eq:diagonalMajoranas}) define how each MBS evolves under the implemented transformation:
\begin{equation}
    \begin{aligned}
        \gamma_1 &\rightarrow \gamma_1^D, && \Tilde{\gamma}_1 \rightarrow \gamma_\Delta^D, \\
        \gamma_2 &\rightarrow \gamma_\theta^{\prime \, D}, && \Tilde{\gamma}_2 \rightarrow \gamma_\eta^D, \\
        \gamma_3 &\rightarrow \gamma_\phi^\prime, && \Tilde{\gamma}_3 \rightarrow \gamma_\eta^\prime.      
    \end{aligned}
\end{equation}
This allows us to express the total parity $\sigma=\pm 1$ at any point of the protocol as
\begin{equation}
    \begin{aligned}\label{eq:totparity}
    \sigma &= i \gamma_1 \Tilde{\gamma_1} \cdot i \gamma_2 \Tilde{\gamma_2} \cdot i \gamma_3 \Tilde{\gamma_3} \\
    &= i \gamma_1^D \gamma_\Delta^D \cdot i \gamma_\theta^{\prime D} \gamma_\eta^D \cdot i \gamma_\phi^\prime \gamma_\eta^\prime.
    \end{aligned}
\end{equation}
Inserting \cref{eq:groundStateCond} into \cref{eq:totparity} yields
\begin{equation}
    i \gamma_\phi^\prime \gamma_\eta^\prime = \sigma \cdot i \gamma_\eta^D \gamma_\theta^{\prime D},
\end{equation}
which we can reinsert into the Hamiltonian to find
\begin{equation}
        \begin{aligned}
        h = \Tilde{\Delta} \, i \gamma_1^D \gamma_\Delta^D
        + (\varepsilon + \sigma \cdot \lambda) \, i \gamma_\eta^D \gamma_\theta^{\prime D}.
    \end{aligned}
\end{equation}
Therefore, the necessary value of $\lambda$ to obtain degeneracy in the ground state sector with total parity $\sigma$ leads to Eq.~(11) in the main paper:
\begin{equation}
\label{eq:degeneracyCondition-SI}
    \varepsilon(\eta, \lambda) + \sigma \lambda = 0.
\end{equation}
In comparison to Eq. (11) of the main paper, \cref{eq:degeneracyCondition-SI} is valid for both total parities $\sigma$. This equation looks simple, but proves to be very difficult to solve in the general case due to the dependence of $\varepsilon$ on $\alpha, \mu$, which in turn depend on $\lambda$. We will show below that for the calculation of the Berry phase it is sufficient to solve this equation during the second protocol step defined by $\rho_1=0$. In this parameter limit it simplifies crucially.

\section{Protocol dependence on parity sector and sign choices}
\cref{eq:degeneracyCondition-SI} shows a crucial difference in the braiding protocol performance between both total parities ($\sigma=\pm1$). Evaluating \cref{eq:degeneracyCondition-SI} for the fermionic limit $(\eta=1)$ at the beginning and end of the protocol $\theta \ll 1$ yields
\begin{equation}
\begin{aligned}
    \varepsilon(\theta, \lambda) &= - \lambda - \theta^2 \frac{1}{\left(-1 + \lambda \right)} + \mathcal{O}(\theta^3), \\
    \Rightarrow  0 =& - \lambda - \theta^2 \frac{1}{\left(-1+ \lambda \right)} + \sigma \, \lambda + \mathcal{O}(\theta^3)  \\ =& (-1+\sigma) \, \lambda -\theta^2 \frac{1}{\left(-1+ \lambda \right)} + \mathcal{O}(\theta^3),
\end{aligned}
\end{equation}
which enforces the condition $\lambda=0+ \mathcal{O}(\theta^2)$ for odd total parity ($\sigma=-1$).
On the other hand, for even total parity ($\sigma=1$), achieving degeneracy requires a big correction $\lambda=1+\mathcal{O}(\theta)$. As a consequence, the low-energy states are degenerate at energies
\begin{equation}
    E_\text{even, low}/\Delta = -1 + \mathcal{O}(\theta),
\end{equation}
but the high-energy states are strongly split into the energies
\begin{equation}
    \begin{aligned}
        E_\text{even, high}/\Delta &= 1 \pm (\varepsilon - \lambda) \\
         &= 1 \pm 2 + \mathcal{O}(\theta).
    \end{aligned}
\end{equation}
Therefore, the correction pulls one of the high-energy states down to $E_\text{even, high}=-1$ and completely closes the gap to the low-energy sector for $\eta = 1$. For $\eta < 1$, the gap remains finite but is reduced compared to the $\sigma = -1$ case in the main paper.

There are therefore two distinct protocols, depending on the total parity. One with a reduced gap and one with a larger gap. Which parity sector that has the larger gap depends on the signs of the different parameters in \cref{eq:hDimless}. To consider the effect of different signs, we check if a sign switch is equivalent to a change of basis of Majoranas, in which case it doesn't affect the resulting physics.

Consider first switching the sign of $\rho_2$. This is equivalent to the transformation 
\begin{equation}
    \begin{aligned}
        \gamma_2 &\rightarrow -\gamma_2\\
        \tilde{\gamma}_2 &\rightarrow -\tilde{\gamma}_2,
    \end{aligned}
\end{equation}
which doesn't change the physics. A similar argument applies to the signs of $\rho_3$ and $\eta$ which also do not affect the protocol. The sign of $\lambda$ is also unimportant since its value is defined by optimizing for degeneracy. 

On the other hand, if $\rho_1$ is negative the protocol will work differently. Changing the sign of $\rho_1$ is equivalent to
\begin{equation}
    \begin{aligned}
        \gamma_1 &\rightarrow - \gamma_1, \\
        \gamma_2 &\rightarrow - \gamma_2, \\
        \gamma_3 &\rightarrow - \gamma_3.
    \end{aligned}
\end{equation}
Since an odd number of Majoranas changed sign, even and odd parity are exchanged. This version of the protocol has a larger gap in the even sector.

Another sign flip is attained with the transformation $\tilde{\gamma}_3 \rightarrow - \tilde{\gamma}_3$ which transforms the Hamiltonian as
\begin{equation}
\begin{aligned}
    \label{eq:hDimless2}
    h \rightarrow \, i \gamma_1 \gamma_\Delta
    &+ \eta \, (\rho_2 \, i \Tilde{\gamma}_1 \Tilde{\gamma}_2 - \rho_3 \, i \Tilde{\gamma}_1 \Tilde{\gamma}_3) \\
    &+ \lambda \, (i \gamma_2 \Tilde{\gamma}_2 - i \gamma_3 \Tilde{\gamma}_3),
\end{aligned}
\end{equation}
and maps even to odd parity. In this form, the gap is larger in the even sector. Note that the error and correction term have the same relative signs between the two terms. If that is violated, the protocol fails to establish degeneracy between the ground states.

For different signs than in the main paper, some of the intermediate formulas in the derivation need to be modified, but the final result for the MBS similarity (Eq.~(15) in the main paper) is the same as long as the error and correction terms have the same relative signs between their respective terms.

\section{Diagonalizing two coupled groups of MBSs}
Analytically diagonalizing the Hamiltonian was possible because the Hamiltonian consists of two groups of MBSs $\vec{\gamma}_L, \vec{\gamma}_R$ with only inter-group couplings. In this case, diagonalization does not mix the two groups, as we show in this section.

The Hamiltonian can be written as
\begin{equation}
    H = i \vec{\gamma}_L^T M \vec{\gamma}_R,
\end{equation}
where $M$ is a real matrix. The singular value decomposition of $M$ is $M=U^TSV$ where $U$ and $V$ are real orthogonal matrices and $S$ is diagonal. Real orthogonal matrices preserve the canonical anticommutation relations, so
\begin{align}
    \vec{\chi}_L &= U\vec{\gamma}_L, \\
    \vec{\chi}_R &= V\vec{\gamma}_R,
\end{align} 
is a valid basis of MBSs. In this basis, the Hamiltonian takes the canonical diagonalized form 
\begin{equation}
    H = i \vec{\chi}_L^T S \vec{\chi}_R,
\end{equation}
which doesn't mix the groups. During adiabatic time-evolution, these groups evolve separately.

\section{Calculation of the Berry phase}
After we diagonalized the Hamiltonian and enforced degeneracy, we can solve the time evolution of the ground state manifold via the calculation of the Berry phase which is given by (Eq.~(12) in the main paper)
\begin{equation}
    U(\eta) = \mathcal{T} \exp{- \oint  [P, \partial_\omega P] \, d \omega},
\end{equation}
i.e., a time-ordered ($\mathcal{T}$) path integral over the full protocol parametrized by $\oint d \omega$. The projection operator onto the ground state manifold $P$ is found by using the parity operator defined in \cref{eq:groundStateCond}
\begin{equation}
    P = \frac{1}{2} (1- i \gamma_1^D \gamma_\Delta^D).
\end{equation}
The solution simplifies drastically if we split the path integral into the three protocol steps defined in the main paper
\begin{equation}
    \oint d \omega = \int\limits_{\pi/2}^0 d \theta + \int\limits_0^{\pi/2} d \phi + \int\limits_0^{\pi/2} d \theta.
\end{equation}
We start by solving the first step of the protocol
\begin{equation}
\label{eq:U21}
    U_{12} = \mathcal{T} \exp{- \int\limits_0^{\pi/2}  [P, \partial_\theta P] \, d \theta}.
\end{equation}

For this, we solve
\begin{equation}
\begin{aligned}
    \partial_\theta P &= -\frac{i}{2} \partial_\theta (\gamma_1^D \gamma_\Delta^D) \\
    &= - \frac{i}{2} (\gamma_\eta^D \gamma_\Delta^D \cdot \partial_\theta \, \theta_\alpha + \gamma_1^D \gamma_\theta^{\prime \, D} \cdot (1+ \partial_\theta \, \theta_\mu) ),
\end{aligned}
\end{equation}
where we made use of
\begin{align}
    \partial_\theta \, \gamma_1^D = \gamma_\eta^D \cdot \partial_\theta \, \theta_\alpha, && \partial_\theta \, \gamma_\Delta^D = \gamma_\theta^{\prime \, D} \cdot (1+ \partial_\theta \, \theta_\mu),
\end{align}
which can be checked via the Eqs. (\cref{eq:deltaMajorana}), (\cref{eq:angleMajoranas}), (\cref{eq:tildeMajoranas}), (\cref{eq:diagonalMajoranas}), and (\cref{eq:angleParametrization}). This yields for the integrand
\begin{equation}
    \begin{aligned}
        [P, \partial_\theta P] = \frac{1}{2} (\gamma_1^D \gamma_\eta^D \cdot \partial_\theta \, \theta_\alpha + \gamma_\Delta^D \gamma_\theta^{\prime \, D} \cdot (1+ \partial_\theta \, \theta_\mu)).
    \end{aligned}
\end{equation}
By applying
\begin{equation}
    \begin{aligned}
        \gamma_1^D \gamma_\eta^D &\stackrel{\phi=0}{=} \gamma_1 \Tilde{\gamma}_2, &&
        \gamma_\Delta^D \gamma_\theta^{\prime \, D} = \gamma_\Delta \gamma_\theta^{\prime} \stackrel{\phi=0}{=} \Tilde{\gamma}_1 \gamma_2,
    \end{aligned}
\end{equation}
we can further simplify the integrand to 
\begin{equation}
\label{eq:integrandT21}
    [P, \partial_\theta P] = \frac{1}{2}( \gamma_1 \Tilde{\gamma}_2  \cdot \partial_\theta \, \theta_\alpha 
                             + \Tilde{\gamma}_1 \gamma_2 \cdot (1+ \partial_\theta \, \theta_\mu) ).
\end{equation}
This simplification was crucial because the integrand of \cref{eq:U21} now consists of two constant matrices which can be commuted as they are acting on different pairs of MBSs. As the integrand commutes with itself throughout the integration, we can neglect the time-ordering $\mathcal{T}$. We find
\begin{equation}
\begin{aligned}
    - \int\limits_0^{\pi/2}  [P, \partial_\theta P] \, d \theta =& \frac{1}{2} \Tilde{\gamma}_2 \gamma_1 \int\limits_0^{\pi/2} \partial_\theta \, \theta_\alpha \, d \theta \\
    &+ \frac{1}{2} \gamma_2 \Tilde{\gamma}_1 \int\limits_0^{\pi/2} (1+\partial_\theta \, \theta_\mu) \, d \theta, \\
    \stackrel{\theta_{\alpha, \mu}|_{\theta=0} =0}{=}& \Tilde{\gamma}_2 \gamma_1 \, \theta_{\alpha,23}/2 + \gamma_2 \Tilde{\gamma}_1 \, (\pi/4 + \theta_{\mu,23}/2).
\end{aligned}
\end{equation}
This solution reveals a particularly interesting fact about the diagonal basis \cref{eq:diagonalMajoranas} parametrized by the angles $\theta_\alpha, \theta_\mu$. These are in general functions of the protocol parametrization $\theta, \phi$ and vary throughout the protocol. However, $\theta_\alpha, \theta_\mu$ are constant during the second protocol step, which can be checked by considering \cref{eq:degeneracyCondition-SI} in the limit $\theta=\pi/2$. This motivates the definition
\begin{align}
\label{eq:halfTimeRot}
    \theta_{\alpha,23} = \theta_\alpha(\theta=\pi/2, \phi), &&  \theta_{\mu,23} = \theta_\mu(\theta=\pi/2, \phi),
\end{align}
describing these angles exactly during the time interval $\tau_{23}$. Using this definition, we obtain the final result
\begin{equation}
    \label{eq:U21final}
    U_{12} = \exp(\Tilde{\gamma}_2 \gamma_1 \, \theta_{\alpha,23}/2 ) \, \exp(\gamma_2 \Tilde{\gamma}_1 \, (\pi/4 + \theta_{\mu,23}/2) ),
\end{equation}
which depends on the previously defined $\theta_{\alpha,23}, \theta_{\mu,23}$ but not on the path they take during the protocol. This is a direct consequence of the path independence of the Berry phase.

This result also provides the Berry phase $U_{31}$ for the third protocol step. We find it by exchanging $2 \rightarrow 3$ and applying the hermitian conjugate to invert the time ordering of the induced coupling
\begin{equation}
    \label{eq:U13final}
    U_{31} = \exp(\gamma_1 \Tilde{\gamma}_3 \, \theta_{\alpha,23}/2 ) \, \exp(\Tilde{\gamma}_1 \gamma_3 \, (\pi/4 + \theta_{\mu,23}/2) ).
\end{equation}
Now the only thing left to solve is the second step of the protocol. As before we begin by calculating
\begin{equation}
    \begin{aligned}
    \label{eq:derivProjT23}
        \partial_\phi P &= -\frac{i}{2} \partial_\phi (\gamma_1^D \gamma_\Delta^D), \\
    &= - \frac{i}{2} ((\partial_\phi \gamma_1^D ) \, \gamma_\Delta^D + \gamma_1^D \, (\partial_\phi \gamma_\Delta^D ) ), \\
    &= - \frac{i}{2} ( \beta_{23} \gamma_\eta^\prime \, \gamma_\Delta^D + \gamma_1^D \, \mu_{23} \gamma_\phi^\prime ) ), \\
    \end{aligned}
\end{equation}
where we defined
\begin{equation}
    \begin{aligned}
        \mu_{23} = \mu(\theta=\pi/2, \phi),
    \end{aligned}
\end{equation}
analogously to \cref{eq:halfTimeRot}. In the same way we define $\nu_{23}, \alpha_{23}, \beta_{23}$. This definition already implies that the basis components are constant during the second protocol step, which was already argued in \cref{eq:halfTimeRot}. This implies
\begin{align}
    \partial_\phi \mu = \partial_\phi \nu = \partial_\phi \alpha = \partial_\phi \beta = 0 && \text{at} && \theta= \pi/2,
\end{align}
which was used as
\begin{equation}
    \begin{aligned}
        \partial_\phi \gamma_1^D &= \beta_{23} \, \partial_\phi \gamma_\eta = \beta_{23} \gamma_\eta^\prime, \\
        \partial_\phi \gamma_\Delta^D &\stackrel{\theta=\pi/2}{=} \mu_{23} \, \partial_\phi \gamma_\Delta \stackrel{\theta=\pi/2}{=} \mu_{23} \, \gamma_\phi^\prime,
    \end{aligned}
\end{equation}
in the calculation of \cref{eq:derivProjT23}. We can use this result to obtain
\begin{equation}
    \begin{aligned}
    \label{eq:integrandT32}
        [P, \partial_\phi P] &= \frac{1}{2} (\beta \, \gamma_1^D \gamma_\eta^\prime +\mu \, \gamma_\Delta^D \gamma_\phi^\prime ) \\
        &= \frac{1}{2} ( \mu^2 \gamma_2 \gamma_3 + \beta^2 \Tilde{\gamma}_2 \Tilde{\gamma}_3 +
        \alpha \beta \gamma_1 \gamma_\eta^\prime - \mu \nu \Tilde{\gamma}_1 \gamma_\phi^\prime ).
    \end{aligned}
\end{equation}
The first two summands of this result are very similar to the integrand obtained in the previous protocol step in \cref{eq:integrandT21}. The difficulty with \cref{eq:integrandT32} lies in the summands proportional to $\gamma_1 \gamma_\eta^\prime$ and $\Tilde{\gamma}_1 \gamma_\phi^\prime$ because these are not constant operators. They evolve during the integration. As a consequence of the operators in the integral of 
\begin{equation}
    \label{eq:U32}
    U_{23} = \mathcal{T} \exp{- \int\limits_0^{\pi/2}  [P, \partial_\phi P] \, d \phi},
\end{equation}
not being constant, also the integrands do not commute with one another. Therefore, we cannot neglect the time ordering $\mathcal{T}$. To solve this expression, we will instead rewrite \cref{eq:U32} into matrix form and afterwards into a differential equation.
We begin by inserting (\cref{eq:angleMajoranas}), (\cref{eq:tildeMajoranas}) and splitting \cref{eq:U32} into two sets of  MBSs
\begin{equation}
\begin{aligned}
    &U_{23} = \\
    & \mathcal{T} \exp{- \frac{1}{2} \int\limits_0^{\pi/2} \mu^2 \gamma_2 \gamma_3 - \mu \nu \Tilde{\gamma}_1 ( - \sin \phi \, \gamma_2 + \cos \phi \, \gamma_3) \, d \phi} \\
    \cdot & \mathcal{T} \exp{- \frac{1}{2} \int\limits_0^{\pi/2} \beta^2 \Tilde{\gamma}_2 \Tilde{\gamma}_3 + 
        \alpha \beta \gamma_1 (-\sin \phi \, \Tilde{\gamma}_2 + \cos \phi \, \Tilde{\gamma}_3) \, d \phi},
\end{aligned}
\end{equation}
which can be solved separately, and we will restrict the presented solution to the first line. This is possible because combinations of $\Tilde{\gamma}_1, \gamma_2, \gamma_3$ commute with combinations of $\gamma_1, \Tilde{\gamma}_2, \Tilde{\gamma}_3$.
\begin{equation}
\begin{aligned}
    \label{eq:U32restricted}
    &U_{23}^{\mu \nu} = \\
    & \mathcal{T} \exp{- \frac{1}{2} \int\limits_0^{\pi/2} \mu^2 \gamma_2 \gamma_3 - \mu \nu \Tilde{\gamma}_1 ( - \sin \phi \, \gamma_2 + \cos \phi \, \gamma_3) \, d \phi}.
\end{aligned}
\end{equation}
Next, we choose a Pauli matrix representation
\begin{align}
    \gamma_3 \Tilde{\gamma}_1 = i \sigma_x, && \gamma_2 \Tilde{\gamma}_1 = -i \sigma_y, && \Rightarrow i \sigma_z = \sigma_x \, \sigma_y = \gamma_2 \gamma_3,
\end{align}
which we insert into \cref{eq:U32restricted}
\begin{equation}
\label{eq:U32restrictedPauli}
    U_{23}^{\mu \nu} = \mathcal{T} \exp{\frac{i}{2} \int\limits_0^{\pi/2} -\mu^2 \sigma_z - \mu \nu (\sin \phi \,  \sigma_y + \cos \phi \, \sigma_x) \, d \phi}.
\end{equation}
Next, we reexpress \cref{eq:U32restrictedPauli} as the solution to a differential equation
\begin{equation}
    \begin{aligned}
        \partial_\omega U_{23}^{\mu \nu} (\omega) = A(\omega) \, U_{23}^{\mu \nu} (\omega),
    \end{aligned}
\end{equation}
evaluated at 
\begin{equation}
    U_{23}^{\mu \nu} = U_{23}^{\mu \nu} (\omega = \pi/2),
\end{equation}
with the initial condition
\begin{equation}
    U_{23}^{\mu \nu} (\omega = 0) = \mathbb{1}_{2},
\end{equation}
where we defined
\begin{equation}
    A(\omega) = \frac{i}{2}(-\mu^2 \sigma_z - \mu \nu (\sin \omega \,  \sigma_y + \cos \omega \, \sigma_x).
\end{equation}
By solving the differential equation, we find
\begin{equation}
    U_{23}^{\mu \nu} = \exp{-i \pi/ 4 \sigma_z} \, \exp{-i \pi/4 \, \nu (\mu \sigma_y - \nu \sigma_z)},
\end{equation}
which we rewrite back into MBSs
\begin{equation}
    U_{23}^{\mu \nu} = \exp{\pi/ 4 \, \gamma_3 \gamma_2} \, \exp{-\pi/4 \, \nu (\mu \gamma_3 \Tilde{\gamma}_1 + \nu \gamma_3 \gamma_2)}.
\end{equation}
Together with the second contribution $U_{23}^{\alpha \beta}$ solved in the same way we find
\begin{equation}
\begin{aligned}
    \label{eq:U32final}
    U_{23} =& \exp{\pi/ 4 \, \gamma_3 \gamma_2} \, \exp{\pi/4 \, \nu (\mu \gamma_3 \Tilde{\gamma}_1 + \nu \gamma_3 \gamma_2)} \\
     &\cdot \exp{\pi/ 4 \, \Tilde \gamma_3 \Tilde \gamma_2} \, \exp{\pi/4 \, \alpha (\beta \Tilde  \gamma_3 \gamma_1 - \alpha \Tilde \gamma_3 \Tilde \gamma_2)}.
\end{aligned}
\end{equation}
The last remaining step is to multiply all contributions to obtain
\begin{equation}
\begin{aligned}
    \label{eq:Ufinal}
    U(\eta) &= U_{31} \, U_{23} \, U_{12} \\
     &= \exp{\pi/4 (1+ \nu_{23}) \gamma_3 \gamma_2} \, \exp{\pi/4 (1- \alpha_{23}) \Tilde \gamma_3 \Tilde \gamma_2},
\end{aligned}
\end{equation}
where we used the identities
\begin{equation}
    \begin{aligned}
        \mu_{23} = \cos (\theta_{\mu,23}), && \nu_{23} = \sin (\theta_{\mu,23}), \\
        \alpha_{23} = \cos (\theta_{\alpha,23}), && \beta_{23} = \sin (\theta_{\alpha,23}). \\
    \end{aligned}
\end{equation}
From this follows that a double braid implements the rotation

\section{Solution of the degeneracy condition in the second protocol interval}
The general solution of \cref{eq:Ufinal} includes the parameters $\nu_{23}, \, \alpha_{23}$ which are obtained by solving the degeneracy condition \cref{eq:degeneracyCondition-SI} in protocol interval $\tau_{23}$ characterized by $\theta=\pi/2$, where \cref{eq:degeneracyCondition-SI} simplifies to 
\begin{equation}
    \begin{aligned}
        \frac{\alpha_{23}}{\mu_{23} } = - \sigma \frac{\lambda_{23}}{\eta },
    \end{aligned}
\end{equation}
which together with
\begin{equation}
    \begin{aligned}
        \frac{\nu_{23}}{\mu_{23} } &= - \tan\left( \frac{1}{2} \arctan\left(\frac{2 \lambda_{23} \eta}{1+\lambda_{23}^2 - \eta^2} \right) \right), \\
        \frac{\beta_{23}}{\alpha_{23}} &= - \tan\left( \arctan\left( - \eta  \frac{\nu_{23}}{\mu_{23}} + \lambda_{23} \right) \right),
    \end{aligned}
\end{equation}
obtained from \cref{eq:thetaSolutions} yields the solution
\begin{equation}
    \label{eq:degeneracySol}
    \begin{aligned}
    \lambda_{23} &= -\sigma\frac{\eta}{\sqrt{1+\eta^2}}, \\
    \nu_{23} &= \frac{\sigma \eta^2}{\sqrt{1+\eta^2 + \eta^4}}, \\
    \alpha_{23} &= \frac{1}{\sqrt{1+\eta^2 + \eta^4}}, \\
    \mu_{23} &= \frac{\sqrt{1+\eta^2}}{\sqrt{1+\eta^2+\eta^4}}, \\
    \beta_{23} &= \frac{\sigma \eta \sqrt{1+\eta^2}}{\sqrt{1+\eta^2 + \eta^4}}.
\end{aligned}
\end{equation}
Eqs.~(\cref{eq:Ufinal}), (\cref{eq:degeneracySol}) together yield the solution for the angles
\begin{equation}
    \begin{aligned}
        \phi =& \frac{\pi}{4} \left( 1+ \nu_{23} \right) 
        =\frac{\pi}{4} \left( 1 +\frac{\sigma \eta^2}{\sqrt{1+\eta^2 + \eta^4}} \right), \\
        \tilde{\phi} =& \frac{\pi}{4} \left( 1 - \alpha_{23} \right) = \frac{\pi}{4} \left( 1 - \frac{1}{\sqrt{1+\eta^2 + \eta^4}} \right),
    \end{aligned}
\end{equation}
presented in Eq.~(13) of the main paper.

\section{Numerical Simulations}
\begin{figure}[t]
    \centering
    \includegraphics[width=1\linewidth]{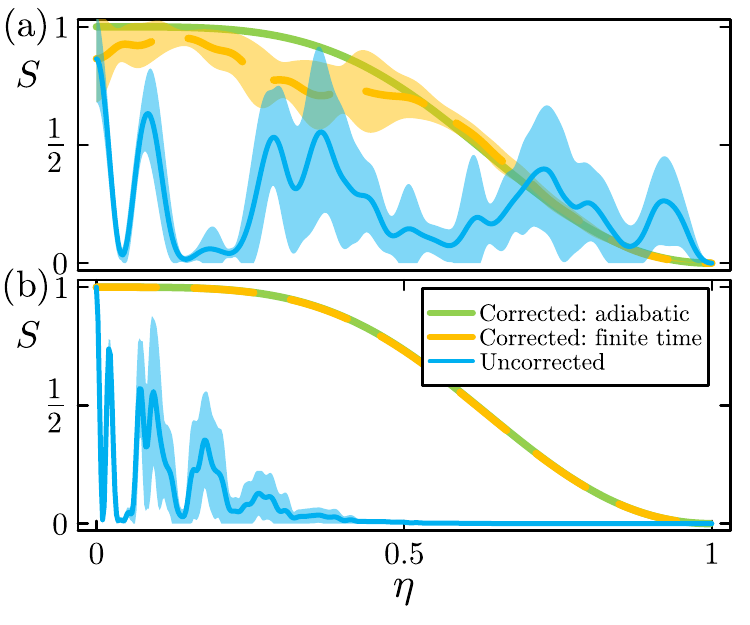}
    \caption{\justifying
    Comparison between finite time numerical simulations of the corrected and uncorrected protocol, and the adiabatic result for the corrected protocol. We show the mean and one standard deviations calculated over a range of protocol times. (a) Short protocol times with $T \Delta = 60 \pm 12, k = 10$.  (b) Longer protocol times with $T \Delta = 200 \pm 40, k = 10$. }
    \label{fig:majorana_similarity_SI}
\end{figure}

To crosscheck our results, investigate diabatic effects and compare with the uncorrected protocol, we solve the time-dependent Schrödinger equation for the full unitary time evolution operator $U(\eta, t)$,
\begin{equation}
\begin{split}
    \partial_t U(\eta,t) &= iH(t)U(\eta,t), \\
    U(\eta,0) &= \mathbb{1},
\end{split}
\end{equation}
with $H(t)$ given by \cref{eq:hDimless} (Eq. (1) of the main paper). The couplings $\rho_i$ are turned on and off periodically according to the function
\begin{equation}
     f(t) = 1 / 2 + \tanh(k \cos(2\pi t/T)) / 2,
\end{equation}
where $k$ determines how sharp the step is and $T$ is the time for a single braid. The different couplings are obtained by shifting the smooth step function as
\begin{equation}
\begin{aligned}
    \rho_1 &= f(t/T)/N, \\
    \rho_2 &= f(t/T - 1/3)/N, \\
    \rho_3 &= f(t/T - 2/3)/N,
\end{aligned}
\end{equation}
where the normalization factor $N$ is there to ensure that $\sum_i \rho_i^2 = 1$. The precise parametrization is unimportant as long as it is adiabatic with respect to the large gap ($T \Delta \gg k$) and diabatic with respect to the smallest gap ($T \Delta e^{-2k} \ll k$). We set $k=10$ in all our simulations. The parameter $\lambda$ is chosen at each point in time so that the ground state sector is degenerate in the appropriate parity sector. This is done by solving \cref{eq:degeneracyCondition-SI} with a numerical root-finding algorithm at each time step. The code we used for these calculations can be found at \cite{nitsch_code_braiding}.

\autoref{fig:majorana_similarity_SI} shows a comparison between numerical results for the MBS similarity and the analytical result of the main paper (Eq.~(15)). For the numerical solutions, we show the mean and one standard deviation by averaging over a range of protocol times $T$. The top figure (a) is for a short protocol time with large diabatic errors. $T$ is sampled uniformly in an interval where $T\Delta = 60 \pm 12$. The bottom figure (b) uses a longer protocol time with $T\Delta = 200 \pm 40$. The corrected protocol has no errors from a dynamical phase, and in this case very small diabatic errors. 

\section{Asymmetric MBS overlaps}
\begin{figure}[t!]
    \centering
    \includegraphics[width=1\linewidth]{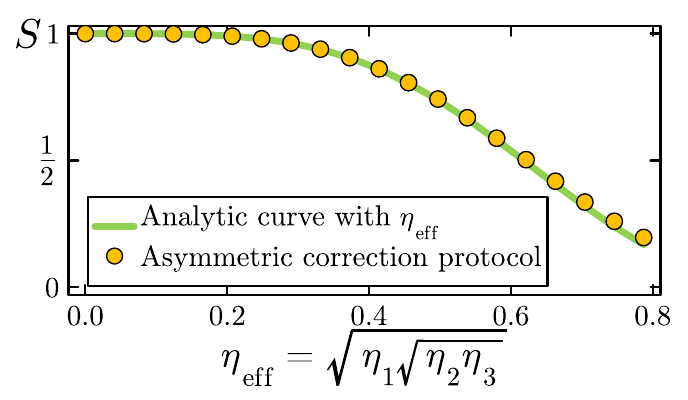}
    \caption{\justifying
    Protocol with asymmetric MBS overlaps $\eta_2 = \eta_1/2, \eta_3 = \eta_1/5$. The markers show the MBS Similarity for numerical solutions with $T\Delta = 2000, k = 10$. The analytic curve is obtained by using $\eta_\mathrm{eff}$ in the analytical formulas. }
    \label{fig:majorana_similarity_asymmetric}
\end{figure}

In the main paper, we have assumed that the MBS overlaps are symmetric, i.e., $\eta_1=\eta_2=\eta_3$. In this section, we consider asymmetric errors and find empirically that a suitably generalized version of the analytical results works very well even in this case. The Hamiltonian is now
\begin{equation}
\begin{aligned}
    H/\Delta = &\rho_1 \, i \gamma_1 \tilde{\gamma}_1 + \rho_2 \, i \gamma_1 \gamma_2 + \rho_3 \, i \gamma_1 \gamma_3 \\
    &+ \rho_2 \sqrt{\eta_1\eta_2} \, i \tilde{\gamma}_1 \tilde{\gamma}_2 + \rho_3 \sqrt{\eta_1\eta_3}\, i \tilde{\gamma}_1 \tilde{\gamma}_3) \\
    &+ \lambda_2\, i \gamma_2 \tilde{\gamma}_2 + \lambda_3\, i \gamma_3 \tilde{\gamma}_3,
\end{aligned}
\end{equation}
and the correction term now includes two parameters: $\lambda_2$ and $\lambda_3$. At each step in time, we numerically optimize these for energy degeneracy of the ground states.

We generalize the analytical result to account for the three different overlaps $\eta_1,\eta_2,\eta_3$ by defining an effective MBS overlap
\begin{equation}
    \eta_\text{eff} = \sqrt{\eta_1\sqrt{\eta_2\eta_3}},
\end{equation}
which we use in in place of $\eta$ in the analytical result. This formula was obtained by testing different ways of taking the mean of the overlaps and this one worked the best. \autoref{fig:majorana_similarity_asymmetric} shows a comparison between a numerical solution and the analytic curve using $\eta_\text{eff}$. We have plotted a range of $\eta_\text{eff}$ where the numerical optimization for $\lambda_2$ and $\lambda_3$ was successful, and the curves agree very well. For larger values of $\eta_\text{eff}$, the optimization failed.

\ifdefined\isarxiv
\else
    \bibliography{main}
\fi

\fi

\end{document}